\documentclass[aps,prd,showkeys,showpacs,amssymb,twocolumn,
amsfonts,epsf,preprintnumbers,nofootinbib,superscriptaddress]{revtex4}

\usepackage[dvips]{graphicx}
\usepackage{bm,latexsym,amsmath,amssymb,amsfonts,color}

\def\be {\begin{equation}}
\def\ee {\end{equation}}
\def\ba {\begin{eqnarray}}
\def\ea {\end{eqnarray}}

\def\c  {\gamma}

\def\bi {\begin{itemize}}
\def\ei {\end{itemize}}
\newcommand\beq{\begin{eqnarray}}
\newcommand\eeq{\end{eqnarray}}
\newcommand{\bea}{\begin{eqnarray}}
\newcommand{\eea}{\end{eqnarray}}

\usepackage{graphicx}
\usepackage{dcolumn}
\usepackage{bm,morefloats,color}

\usepackage[
      colorlinks=true,
      linkcolor=blue,
      urlcolor=blue,
      filecolor=blue,
      citecolor=red,
      pdfstartview=FitV,
      pdftitle={},
      pdfauthor={},
      pdfsubject={},
      pdfkeywords={},
      pdfpagemode=None,
      bookmarksopen=true
]{hyperref}

\usepackage{amssymb,bm}

\def\X5sp{{\rm X}_5}
\def\Y3sp{{\rm Y}_3}
\def\Z3sp{{\rm Z}_3}

\begin{document}

\title{Unified approach to the entropy of an extremal
rotating BTZ black hole: Thin shells and
horizon limits}
\author{Jos\'e P. S. Lemos}
\email{joselemos@ist.utl.pt}
\affiliation{Centro de Astrof\'isica e Gravita\c c\~ao - CENTRA,
Departamento de F\'{\i}sica, Instituto Superior T\'ecnico - IST,
Universidade de Lisboa - UL, Avenida Rovisco Pais 1, 1049-001,
Portugal, and \\
Gravitational Physics Group,
Faculty of Physics, University of Vienna, Boltzmanngasse 5,
A1090, Wien, Austria}
\author{Masato Minamitsuji}
\email{masato.minamitsuji@ist.utl.pt}
\affiliation{Centro de Astrof\'isica e Gravita\c c\~ao - CENTRA,
Departamento de F\'{\i}sica, Instituto Superior T\'ecnico - IST,
Universidade de Lisboa - UL, Avenida Rovisco Pais 1, 1049-001,
Portugal.}
\author{Oleg B. Zaslavskii}
\email{zaslav@ukr.net}
\affiliation{Department of Physics and Technology, Kharkov V. N. Karazin National
University, 4 Svoboda Square, Kharkov 61022, Ukraine, and Institute of
Mathematics and Mechanics, Kazan Federal University, 18 Kremlyovskaya St.,
Kazan 420008, Russia.}

\begin{abstract}

Using a thin shell, the first law of thermodynamics, and a unified
approach, we study the thermodymanics and find the entropy of a
(2+1)-dimensional extremal rotating Ba\~{n}ados-Teitelbom-Zanelli
(BTZ) black hole.  The shell in (2+1) dimensions, i.e., a ring, is
taken to be circularly symmetric and rotating, with the inner region
being a ground state of the anti-de Sitter (AdS) spacetime and the
outer region being the rotating BTZ spacetime.  The extremal BTZ
rotating black hole can be obtained in three different ways depending
on the way the shell approaches its own gravitational or horizon
radius.  These ways are explicitly worked out.  The resulting three
cases give that the BTZ black hole entropy is either the
Bekenstein-Hawking entropy, $S=\frac{A_+}{4G}$, or it is an arbitrary
function of $A_+$, $S=S(A_+)$, where $A_+=2\pi r_+$ is the area, i.e.,
the perimeter, of the event horizon in (2+1) dimensions.  We speculate
that the entropy of an extremal black hole should obey $0\leq S(A_+)
\leq\frac{A_+}{4G}$. We also show that the contributions from the
various thermodynamic quantities, namely, the mass, the circular
velocity, and the temperature, for the entropy in all three cases are
distinct. This study complements the previous studies in thin shell
thermodynamics and entropy for BTZ black holes.  It also corroborates
the results found for a (3+1)-dimensional extremal electrically
charged Reissner-Nordstr\"om black hole.

\end{abstract}

\pacs{
04.70.Dy, 04.40.Nr. 
}
\keywords{Quantum aspects of black holes, thermodynamics, three-dimensional
black holes, spacetimes with fluids}
\date{\today}
\maketitle

\maketitle

\section{Introduction}
\label{sec1}

One can argue that thin matter shells in general relativity provide
the simplest class of spacetimes after vacuum spacetimes. Indeed, thin
shells, besides giving instances of static and dynamic spacetimes,
allow themselves to be scrutinized in relation to their entropic and
thermodynamic matter and gravitational properties, and even from
those  properties
to pick up the corresponding black hole properties. For static and
rotating circularly symmetric thin shells, i.e., thin rings, in
(2+1)-dimensional  Ba\~{n}ados-Teitelbom-Zanelli
(BTZ) spacetimes their entropic and thermodynamic
properties have been worked out in general and in the limit where the
ring is taken to its own gravitational, or horizon, radius, i.e., in the black hole
limit \cite{quintalemosbtzshell,energycond,btzshell,extremalbtz}.  For
static electric charged spherically symmetric thin shells in 
(3+1)-dimensional Reissner-Nordstr\"om spacetimes these properties have
also been worked out in detail in general and in the black hole limit
\cite{charged,extremalshell,lqzn}, see also \cite{martin} 
for neutral thin shells
in Schwarzschild spacetimes.
Related studies, where the entropy of black holes can
be studied through systems with matter, involve
quasiblack holes for which
matter is spread over a
3-dimensional spatial region rather than on a 2-dimensional thin
shell \cite{quasi_bh1,quasi_bh2}, or are connected
to a quasistatic collapse of
matter \cite{pretisrvol}.
These works
\cite{quintalemosbtzshell,energycond,btzshell,extremalbtz,charged,extremalshell,lqzn,martin,quasi_bh1,quasi_bh2,pretisrvol}
stem from the fact that the concept of entropy is originally based on
quantum properties of matter, and so it is very important to study
whether and how black hole thermodynamics could emerge from
thermodynamics of collapsing matter, when matter is compressed within
its own gravitational radius.
Conversely, it is through 
black hole entropy that we can grasp the microscopic aspects of a spacetime
and hence of quantum gravity, 
and the
fact that thermodynamics of a thin shell reflects thermodynamic
properties of a black hole formed after quasistatic
collapse of the shell indicates some connection between matter
and gravitational degrees of freedom.

In this thin shell approach to black hole
entropy a clear cut distinction exists between nonextremal black holes
and extremal black holes.

For nonextremal black holes one finds that the entropy is
\begin{equation}
S=\frac{A_+}{4G}\,,
\label{ent1bh}
\end{equation}
where $A_{+}$ is the area of the event horizon and
$G$ is the gravitational constant. Throughout the paper, we use units
such that the velocity of the light, the Planck constant, and the
Boltzmann constant are set to 1. This result
has been found for static BTZ shells 
\cite{quintalemosbtzshell} and for rotating BTZ shells 
\cite{energycond,btzshell},
as well as for Reissner-Nordstr\"om shells
\cite{charged},
all in the black hole limit.  The result
recovers the Bekenstein-Hawking entropy formula
in (2+1) dimensions \cite{btz,carlip}
and in the original works in
(3+1) dimensions \cite{bek1,bch,haw}.
In (2+1) dimensions, $A_+$ is a perimeter $A_+=2\pi r_+$,
and in (3+1) dimensions, $A_+=4\pi r_+^2$ is the usual
area, with $r_+$ being the gravitational or horizon
radius.

For extremal black holes,  the ones 
which we will study in this paper,
the situation is more
subtle in the shell approach. Extremal
black holes
are those for which the angular momentum or electric charge is equal to the mass
in some appropriate units.
It has been found that
the entropy of the extremal black hole can depend
on the way the shell approaches its own gravitational radius.
This results in three cases.
On one hand, clearly, there is a case
for an originally nonextremal shell, which we call {\it Case 1}, in which
after taking the black hole limit the shell 
turns into an extremal shell, where  one finds
$S=\frac{A_+}{4G}$ as in Eq.~(\ref{ent1bh}), 
see also \cite{btzshell,extremalbtz} for BTZ and 
\cite{charged,extremalshell,lqzn} for Reissner-Nordstr\"om.
On the other hand, it was further found in
the Reissner-Nordstr\"om situation
that there is a new case \cite{lqzn}, which we call {\it Case 2},
in which the shell is turned extremal concomitantly
with the spacetime being turned into a black hole.
In this case one finds also $S=\frac{A_+}{4G}$ as in Eq.~(\ref{ent1bh}).
Finally, for an ab initio extremal shell that turns into an extremal black
hole, one finds that the entropy is a generic function of $A_+$, i.e., 
\begin{equation}
S=S(A_+)\,.
\label{ente11bh}
\end{equation}
This result, which we call {\it Case 3},
is found 
both in extremal rotating BTZ \cite{extremalbtz} and in extremal
electric charged Reissner-Nordstr\"om \cite{extremalshell}.

Given the result (\ref{ente11bh}) together with
(\ref{ent1bh}) one is led to speculate that
the entropy of an extremal black hole should obey
\begin{equation}
0\leq S(A_+) \leq\frac{A_+}{4G}\,.
\label{lowupent1bh}
\end{equation}
The lower limit  
\begin{equation}
S=0\,,
\label{ent1ebh}
\end{equation}
is indeed found through a Euclidean path integral
approach to extremal black hole entropy, both
in  BTZ black holes \cite{ebh2}
and in Reissner-Nordstr\"om
black holes \cite{ebh1}, 
whereas in contradiction,
the Bekenstein-Hawking upper limit of Eq.~(\ref{lowupent1bh}),
$S=\frac{A_+}{4G}$, see also Eq.~(\ref{ent1bh}), 
is 
found through string theory techniques
in extremal cases,
namely, in
(2+1)-dimensional extremal 
rotating BTZ black holes 
\cite{birmsacsen}, and in 
(3+1)-dimensional extremal
Reissner-Nordstr\"om black holes
\cite{string11},  following the breakthrough
worked out in (4+1) dimensions \cite{string1,string2},
see also \cite{ebh3,ebh4,ebh5,ebh6,ebh7,ghoshmitra,string3,string4,string5,cano1}
for further studies on thermodynamics and entropy of extremal black holes. 
In a sense, Eq.~(\ref{lowupent1bh}) fills the gap
between Euclidean path integral
approaches and string theory techniques for the
entropy of extremal black holes.

The aim of this paper is to complete the study on extremal
rotating BTZ thin shell
thermodynamics \cite{quintalemosbtzshell,energycond,btzshell,extremalbtz},
in order to have a
full understanding of the entropy of an
extremal rotating BTZ black hole.
We follow
also the studies for electrically charged  Reissner-Nordstr\"om
shells
\cite{charged,extremalshell}
and in particular we adopt
the unified approach devised
for an electrically
charged Reissner-Nordstr\"om
thin shell \cite{lqzn}, and
study the three different limits of a rotating thin shell in a
(2+1)-dimensional rotating BTZ spacetime when it 
approaches both extremality and its own gravitational radius,
i.e., in the extremal BTZ black hole limit.
These three different limits
yield the three cases, {\it Cases 1-3}, 
already mentioned.
Our analysis will point out the similarities
between the rotating and the electric charged case
and will show the contributions from the various
thermodynamic quantities appearing in the first law to the entropy in
all three cases.
The approach developed in the present work can be of interest
for the generic investigation of black hole entropy in the thin shell formalism,
in particular,  
for the Kerr black hole, at least in the slow
rotation approximation, 
or to
other more complicated (3+1)-
and ($n$+1)-dimensional black holes, with $n>3$.

The paper is organized as follows.
In Sec.~\ref{sec2}, we review the
mechanics and thermodynamics of a rotating thin shell in (2+1)
dimensions with a negative cosmological constant, where the exterior
spacetime is BTZ.  In Sec.~\ref{sec3}, we introduce the three
different limits, thus establishing
three different cases, when the rotating thin shell
is taken into its own gravitational radius, and forms
an extremal BTZ black
hole.  We define the good variables to study these limits, and work out
the geometry, the mass, and the angular momentum of the shell
in the three different cases.  In
Sec.~\ref{sec4}, we discuss the three different cases for the
pressure, the circular velocity,  and the local temperature of the shell.
In Sec.~\ref{sec5}, we calculate the
entropy of a rotating extremal BTZ black hole in the three different
cases.  In Sec.~\ref{sec6}, we show, in the three different cases,
which terms in the first law give the dominant contributions to the
entropy.  In Sec.~\ref{sec7}, we conclude.

\section{Thin shell thermodynamics in a (2+1)-dimensional BTZ  
spacetime\label{sec2}}

We consider general relativity in (2+1) dimensions with a cosmological constant $\Lambda$, 
where we assume that $\Lambda<0$, so that the spacetime is asymptotically
anti-de Sitter (AdS), with
curvature scale $\ell=\sqrt{-\frac{1}{\Lambda}}$.
In an otherwise vacuum spacetime,
we introduce a timelike rotating thin shell, i.e., a timelike thin ring in the
(2+1)-dimensional spacetime, with radius $R$, that divides the
spacetime into the inner and outer  regions. 
We leave $G$ explicitly in the formulas, the other physical constants
are set to 1.

The spacetime inside the shell, $0<r<R$, where  $r$ is a radial coordinate,
is given by the zero mass $m=0$ BTZ-AdS solution in (2+1) dimensions.

The spacetime outside the shell, $r>R$,
is generically
described by the rotating BTZ solution with
Arnowitt-Deser-Misner (ADM) mass $m$
and angular momentum $\cal J$.
Two important quantities of the outer spacetime,
which are related to $m$ and ${\cal J}$,
are the gravitational radius $r_+$
and the Cauchy radius $r_-$. The relations between the quantities
are \cite{btz}
\bea
\label{ml1}
8G\ell^2
m=r_+^2+ r_-^2\,,
\eea
\bea
\label{jl1}
4G\ell {\cal J}=r_+r_-\,.
\eea 
From Eqs.~(\ref{ml1})
and~(\ref{jl1}),
one clearly sees that
one can trade $m$ and $\cal J$ for $r_+$ and $r_-$ and vice versa.
For a spacetime that is not over rotating, as will be the case
considered here, one has that $m\geq\frac{\cal J}{\ell}$
which, in terms of $r_+$
and $r_-$, translates into $r_+\geq r_-$.
This inequality is
saturated in the extremal case, $r_+=r_-$, i.e., $m=\frac{{\cal J}}{\ell}$.
The gravitational area $A_+$ defined as
\bea
A_+=2\pi r_+\,
\label{areah}
\eea
is actually a perimeter, since there are just 2 space dimensions.

The shell itself has radius $R$, and it is
quasistatic in the sense that
$\frac{dR}{d\tau}=\frac{d^2R}{d\tau^2}=0$,
where $\tau$ is the proper time on the shell.
The area $A$ of the shell defined as 
\bea
A=2\pi R\,
\label{areashell}
\eea
is also a perimeter, since there are 2 space
dimensions.
We assume that the shell
is always located outside or at the gravitational radius,
\bea
R\geq r_+\,.
\label{Rgeqr+}
\eea
Note that the gravitational radius is not a horizon radius
in this case, it is simply a feature of
the spacetime.
It would be a horizon radius only
if $R\leq r_+$.
Since from Eq.~(\ref{Rgeqr+}) one has $R\geq r_+$,
there is a horizon only in the limit $R=r_+$. In this limiting
situation the shell is on the verge of becoming a black hole.
Besides having a radius $R$, the shell has mass
$M$ and angular momentum $J$.

To find the properties of the shell and the connection to the
inner and outer spacetime one has to work out
the junction conditions. 
The junction conditions determine
the energy density of the shell $\sigma$
and the angular momentum density of the shell $j$,
or if one prefers, the rest mass of the
shell $M\equiv2\pi R\sigma$
and the 
angular momentum of the shell
$J\equiv2\pi R$. One finds that
$M$ and $J$ 
are some specific functions of the 
ADM spacetime mass $m$, 
angular momentum $\cal J$, and the shell's radius $R$,
see 
\cite{btzshell} for details (see also \cite{energycond}).
These relations can be inverted to give
the ADM spacetime mass $m$ as a function
of $M$, $J$ and $R$, namely,
\bea
m(M,J,R)
=
 \frac{R M}{\ell}
-2GM^2
+\frac{2G}{R^2}J^2\,,
\label{mMJ}
\eea
and the ADM spacetime angular momentum  $\cal J$ also as a function
of $M$, $J$ and $R$, namely,
\bea
{\cal J}(M,J,R)=J\,.
\label{JJ}
\eea
In Eqs.~(\ref{mMJ}) and~(\ref{JJ})
we have written $m$ as $m(M,J,R)$
and  $\cal J$ as ${\cal J}(M,J,R)$
in order to
make manifest the explicit dependence of the ADM spacetime
mass $m$ and the ADM spacetime
angular momentum $\cal J$ 
on the shell quantities, i.e., its rest mass $M$,
its angular momentum $J$, and its radius $R$.
This explicit dependence is also useful when we deal with
the thermodynamics of the shell.
The gravitational radius $r_+$ and the Cauchy
radius $r_-$ can be found inverting
Eqs.~(\ref{ml1}) and~(\ref{jl1}) \cite{btz}. 
The gravitational radius
is  
\bea
r_+(M,J,R)=
2\ell
\sqrt{
Gm
+\sqrt{(Gm)^2-\frac{(G{\cal J})^2}{\ell^2}}}
\,,
\label{ml}
\eea
and the Cauchy radius is 
\bea
r_-(M,J,R)=
2\ell
\sqrt{
Gm
-\sqrt{(Gm)^2-\frac{(G{\cal J})^2}{\ell^2}}
}
\,,
\label{jl}
\eea
with $m$ and $\cal J$ seen as functions of
$M$, $J$, and $R$ through Eqs.~(\ref{mMJ})
and~(\ref{JJ}).

As a thermodynamic system,
 the shell has a
locally measured temperature $T$ and
an entropy $S$. We consider that the shell is
adiabatic, i.e., it does not radiate to the
exterior.
The entropy $S$ of a system can be expressed
as a function of the state independent variables
which for the rotating shell 
can be chosen as
the shell's locally measured proper mass $M$, 
the shell's angular momentum $J$, and the shell's area $A$.
Thus, $S=S(M,J,A)$ and in these variables the
first law of thermodynamics reads
\bea
\label{1st}
TdS=dM +p\, d A-\Omega\, dJ\,,
\eea
where $p$ is the tangential pressure at the shell,
$\Omega$ is 
the thermodynamic angular velocity of the
shell, and $T$ is 
the temperature of the shell. These quantities,
$p$, $\Omega$, and $T$ 
are equations of state functions of
$(M,J,A)$, i.e.
$p=p(M,J,A)$, $\Omega=\Omega(M,J,A)$, and $T=T(M,J,A)$.
In (2+1) dimensions, the shell's area is a perimeter, namely,
$A=2\pi R$,
and we can thus express $S$, $p$, $\Omega$, and $T$, as  functions of
the shell radius $R$,
instead of its area $A$. This simplifies the
presentation. Thus,
$S=S(M,J,R)$, $T=T(M,J,R)$, $p=p(M,J,R)$, and $\Omega=\Omega(M,J,R)$.
In order to have a well-defined entropy $S$
there are integrability conditions for
$T=T(M,J,R)$, $p=p(M,J,R)$, and $\Omega=\Omega(M,J,R)$, see
\cite{btzshell}.

The first law for the shell, Eq.~(\ref{1st}),
is clearly displayed and has a clear
physical meaning in the variables $M$, $J$, and $R$.  As it turns out
and as we will see, it is much simpler mathematically to work instead
in the variables $r_+$, $r_-$, and $R$.  Indeed, from Eqs.~(\ref{ml})
and~(\ref{jl}) together with Eqs.~(\ref{mMJ}) and~(\ref{JJ}), one can
swap the variables $M$, $R$, and $J$, into $r_+$, $r_-$, and $R$.  So,
from now on, we express our quantities in terms of $(r_+,r_-,R)$.

Inverting  Eq.~(\ref{mMJ}) and using  Eqs.~(\ref{ml})
and~(\ref{jl}) together with Eq.~(\ref{JJ}),
one finds
\bea
\hskip -0.4cm M(r_+,r_-,R)
=
\frac{R}{4 G\ell}
\Big(
1-\frac{1}{R^2}\sqrt{(R^2-r_+^2)(R^2-r_-^2)}
\Big)
\label{sig1}.
\eea

Inverting  Eq.~(\ref{JJ}) and using  Eqs.~(\ref{ml})
and~(\ref{jl}) (or more simply Eq.~(\ref{ml1})),  one finds
\bea
J(r_+,r_-,R)=\frac{r_+r_-}{4G \ell }\,.
\label{j1}
\eea

The tangential pressure $p$ at the shell
found through the junction conditions 
\cite{btzshell} (see also \cite{energycond})
is 
\bea
p(r_+,r_-,R)
=
\frac{1}{8\pi G \ell}
\Bigg(
\frac{R^4-r_+^2 r_-^2}{R^2\sqrt{(R^2-r_+^2)(R^2-r_-^2)}}-1
\Bigg)\,.
\nonumber 
\\
\label{ne1}
\eea

The angular velocity $\Omega$ and the corresponding linear or
circular velocity $v=R\,\Omega$ can be found either by 
the junction conditions or from one integrability
condition of the first law of thermodynamics Eq.~(\ref{1st})
\cite{btzshell,extremalbtz}. 
The integrability condition gives that 
the angular velocity defined thermodynamically can be expressed by
$\Omega(r_+,r_-,R)=\frac{ r_+r_-}{R
\sqrt{
\left(1-\frac{r_+^2}{R^2}\right)
\left(1-\frac{r_-^2}{R^2}\right)}}
\Big(
c(r_+,r_-)-\frac{1}{R^2}
\Big)$,
where $c(r_+,r_-)$ is an integrating arbitrary function of $r_+$ and $r_-$
(see  Eq.~(59) in \cite{btzshell} 
and Sec.~VI in
\cite{extremalbtz}).
We choose 
$
c(r_+,r_-)=\frac{1}{r_+^2},
$
in order to have the well-defined black hole limit \cite{btzshell,extremalbtz}.  
In this case, one sees that $\Omega$ vanishes 
when the shell approaches the gravitational radius, 
$R\to r_+$.
Since the circular velocity of the shell is
$v=R\,\Omega$, one has,
with the choice $c(r_+,r_-)=\frac{1}{r_+^2}$
and  
after simplifications, that
\bea
\label{ne3}
v\big(r_+,r_-,R\big)=R\,\Omega(r_+,r_-,R)=
\frac{r_-}{r_+}
\sqrt{\frac{R^2-r_+^2}{R^2-r_-^2}}\,.
\eea

The temperature $T$ being a pure
thermodynamic quantity is found from another integrability
condition of the first law of thermodynamics Eq.~(\ref{1st})
\cite{btzshell,extremalbtz}. 
As found in \cite{btzshell}, the temperature can be expressed as
$
T(r_+,r_-,R)=
\frac{T_0(r_+,r_-)}{
\frac{R}{\ell}
\sqrt{
\left(1-\frac{r_+^2}{R^2}\right)
\left(1-\frac{r_-^2}{R^2}\right)}
}
$,
where $T_0(r_+,r_-)$ is an arbitrary function of $r_+$ and $r_-$
(see also Eqs.~(C2) and~(C3) from \cite{extremalbtz}).
Now, $T_0(r_+,r_-)$ is chosen
to be the Hawking temperature of the BTZ black hole, i.e., 
$
T_0(r_+,r_-)= T_H (r_+,r_-)=\frac{1}{2\pi \ell^2}\frac{r_+^2-r_-^2}{r_+}$
\cite{btz}. Thus, we have
\bea
T\big(r_+,r_-,R\big)
=\frac{r_+^2-r_-^2}{2\pi \ell R  r_+} 
\frac{R^2}
       {\sqrt{\big(R^2-r_+^2\big)\big(R^2-r_-^2\big)}}.
\label{ne2}       
\eea
For the outer spacetime, it is usually useful
to define the redshift function $k$
that appears naturally in several instances,
namely, 
$
k\big(r_+,r_-,R\big)
=\frac{R}{\ell}
\sqrt{
\left(1-\frac{r_+^2}{R^2}\right)
\left(1-\frac{r_-^2}{R^2}\right)}
$.
With this quantity, the temperature $T$ assumes
the familiar form
$
T(r_+,r_-,R)=
\frac{T_H(r_+,r_-)}{k (r_+,r_-,R)}
$, and so 
the function $T_H(r_+,r_-)$
can be interpreted as the temperature of the shell located at
the radius where $k=1$, the Hawking temperature.
Seen in this fashion, the formula for $T$ expresses then the
gravitational redshift of the temperature of the shell,
namely,
it is an instance of the Tolman temperature formula.

Note that
the choices, $
c(r_+,r_-)=\frac{1}{r_+^2}
$ for the velocity $v$
and
$
T_0(r_+,r_-)= T_H (r_+,r_-)=\frac{1}{2\pi \ell^2}\frac{r_+^2-r_-^2}{r_+}
$ for the temperature $T$
that lead to Eqs.~\eqref{ne3} and \eqref{ne2},
respectively,
are essential if we want to take the black hole limit, i.e.,
when the shell is
taken to its gravitational radius, $R\to r_+$
\cite{btzshell,extremalbtz}.
So we stick to these choices.

\section{The three different approaches and the three limits
to the BTZ extremal black hole} \label{sec3}

\subsection{The variables useful to
define the three different
approaches and limits to an extremal horizon}

To study the entropy of the BTZ extremal black hole
we take a unified
approach, see \cite{lqzn} for an extremal electric
charged shell in 3+1 dimensions. 
For that, we introduce the dimensionless
parameters $\varepsilon$ and $\delta$ through
\begin{equation}
\varepsilon ^{2}=1-\frac{r_{+}^2}{R^2}\,,  \label{e}
\end{equation}
\begin{equation}
\delta ^{2}=1-\frac{r_{-}^2}{R^2}\,. \label{d}
\end{equation}

From Eqs.~(\ref{e}) and (\ref{d}), we see
that we can change
the independent thermodynamics
variables $(r_+,r_-,R)$ into
the new 
variables
$(\varepsilon,\delta,R)$.
In this set of variables, for example,
the redshift function $k$ defined above
takes the simple form 
$
k(\varepsilon,\delta,R)=\frac{R}{\ell}
\varepsilon \delta
$.

\subsection{Geometry and the three horizon limits}

The three relevant limits to an extremal black hole
are:

\vskip 0.3cm \noindent \textit{Case 1:} 
$r_+\neq r_-$ and 
$R\to r_+$, i.e., 
\begin{equation}
\delta ={O}(1)\,,\quad\varepsilon \to 0\,.  \label{de1}
\end{equation}
In evaluating the entropy $S$, we then take
$r_+\to r_-$, i.e., the $\delta\to0$ limit, to make the shell  extremal
at its
own gravitational radius $R=r_+$.

\vskip 0.3cm \noindent \textit{Case 2:} 
$r_{+}\rightarrow r_{-}$ and $R\rightarrow
r_{+}$, i.e., \noindent 
\begin{equation}
\delta =\frac{\varepsilon }{\lambda}\,,\quad\varepsilon \to 0\,,  \label{de}
\end{equation}
where the constant $\lambda$
is finite, not infinitesimal, and must satisfy $\lambda < 1$
due to $r_{+}> r_{-}$. 
The limit $\varepsilon \rightarrow 0$ means here that
simultaneously $R\rightarrow r_{+}$ and $r_{+}\rightarrow r_{-}$ in such a
way that $\delta \sim \varepsilon$. 
In other words, extremality and black holeness are approached
concomitantly.

\vskip 0.3cm \noindent \textit{Case 3:} 
$r_+=r_-$ and $R\to r_+$, i.e., 
\begin{equation}
\delta=\varepsilon \,,\quad
\varepsilon \to 0\,.  \label{de2}
\end{equation}
This is the case in which there is an extremal shell from the
very beginning 
and then one pushes it to its own gravitational radius.

\subsection{Mass and angular momentum in the three horizon limits}

In the variables $\varepsilon$ and $\delta$ of Eqs.~(\ref{e}) and (\ref{d}),
the shell's rest mass $M$ in Eq.~\eqref{sig1} can be written as
\begin{equation}
M(\varepsilon,\delta,R)=
\frac{R}{4G\ell}
(1-\varepsilon \delta )\,.  \label{med}
\end{equation}
In addition, from Eq.~\eqref{j1},
the shell's angular momentum $J$ is now 
\begin{equation}
J(\varepsilon,\delta,R)=
\frac{R^2}{4G\ell}\sqrt{(1-\varepsilon ^{2})(1-\delta ^{2})}\,.
\label{jed}
\end{equation}
In all \textit{Cases} 1-3,
the limits defined in Eqs.~(\ref{de1})-(\ref{de2})
yield
\begin{equation}
M(\varepsilon,\delta,r_+)= \frac{r_+}{4G\ell}
\end{equation}
and
\begin{equation}
J(\varepsilon,\delta,r_+)= \frac{r_+^2}{4G\ell} \,,
\label{mqk1}
\end{equation}
for the shell's mass and angular momentum, respectively. 
Thus, the three limits, not surprisingly, yield the same
extremal condition, 
\begin{equation}
J=r_+ M\,.
\end{equation}

\section{Pressure, circular velocity,
and local temperature:
The three extremal BTZ black hole limits} \label{sec4} 

\subsection{Pressure in the three horizon limits}
\label{pphit}

In the variables $\varepsilon$ and $\delta$ of Eqs.~(\ref{e}) and (\ref{d}),
the shell's pressure $p$ in Eq.~\eqref{ne1} can be written as
\begin{equation}
p(\varepsilon,\delta,R) =\frac{1}{8\pi G\ell}
\Big(
\frac{\delta}{\varepsilon}+\frac{\varepsilon}{\delta}
-1-\varepsilon\delta
\Big)
\,.  \label{pd}
\end{equation}
For the \textit{Cases} 1-3,
the limits defined in Eqs.~(\ref{de1})-(\ref{de2})
yield from Eq.~\eqref{pd} 
the expressions
for the pressure as below.

\vskip 0.3cm \noindent \textit{Case1:}
For $\delta ={O}(1)$ and $\varepsilon \to 0$,
\begin{equation}
p(\varepsilon,\delta,r_+) =
\frac{\delta }{8\pi  G \ell\,\varepsilon }\,,  \label{pdiv}
\end{equation}
up to leading order.
Equation~(\ref{pdiv}) means that the pressure is divergent as $1/\varepsilon$.

\vskip 0.3cm \noindent \textit{Case 2:} For $\delta =\frac{\varepsilon }{\lambda}$
and $\varepsilon \to0 $,
\begin{equation}
p(\varepsilon,\delta,r_+) = \frac{1}{8\pi G \ell}
\Big(
\frac{1}{\lambda}+\lambda-1
\Big)\,,  \label{p3}
\end{equation}
up to leading order. Equation~(\ref{p3}) means that the pressure remains
finite but nonzero, since $\lambda$ is finite and fixed with
$\lambda<1$.

\vskip 0.3cm \noindent \textit{Case 3:} For $\delta=\varepsilon$ and $\varepsilon \to 0$,
\begin{equation}
p(\varepsilon,\delta,r_+)= \frac{1}{8\pi G\ell}\,,  \label{pe}
\end{equation}
up to leading order. Equation~(\ref{pe}) means that the pressure remains finite and nonzero.
Note the difference from the (3+1)-dimensional electric
extreme shell in an
asymptotically flat spacetime studied in \cite{extremalshell,lqzn}, where in this same limit
one found instead $p=0$.
This difference arises from the different asymptotic behaviors of the spacetime,
namely, asymptotically flat spacetime in \cite{extremalshell,lqzn}
and 
an asymptotically AdS spacetime here, see also \cite{extremalbtz}.

\subsection{Circular velocity  in the three horizon limits}

With the variables $\varepsilon$ and $\delta$ defined in Eqs.~(\ref{e}) and (\ref{d}),
the shell's circular velocity $v$ in Eq.~\eqref{ne3} can be written as
\begin{equation}
v (R,\varepsilon ,\delta )=\sqrt{\frac{1-\delta ^{2}}{1-\varepsilon ^{2}}}
\,\,\frac{\varepsilon }{\delta }\,.  \label{cd}
\end{equation}
For \textit{Cases} 1-3,
the limits defined in Eqs.~(\ref{de1})-(\ref{de2})
yield from Eq.~\eqref{cd} 
the expressions
for the circular velocity as below.

\vskip 0.3cm \noindent {\it Case 1:} For $\delta ={O}(1)$ and $\varepsilon \to 0$,
\begin{equation}
v(\varepsilon,\delta,r_+)=0\,
\end{equation}
up to leading order.

\vskip 0.3cm \noindent {\it Case 2:} For $\delta =\frac{\varepsilon }{\lambda}$
and $\varepsilon \to0 $,
\begin{equation}
\label{v1}
v(\varepsilon,\delta,r_+)= \lambda \,,
\end{equation}
up to leading order. 
Equation~(\ref{v1}) means that the circular velocity is nonzero since
$\lambda$ is finite and fixed with
$\lambda<1$.

\vskip0.3cm \noindent \textit{Case 3:} For
$\delta =\varepsilon $ and $\varepsilon \rightarrow 0$,
\begin{equation}
v (\varepsilon ,\delta,r_{+})\leq 1\,.  \label{f2}
\end{equation}
This result is not found directly from Eq.~(\ref{cd}).  Indeed, from
Eq.~(\ref{cd}) it follows that $v(r_{+},\varepsilon,\delta)=1$.
However, in this case, the condition $c(r_+r_-)=1/r_+^2$ imposed to
obtain Eq.~(\ref{ne3}) is no longer valid. An independent calculation
is requested for an ab initio extremal shell as shown in
\cite{extremalbtz}.  In this case, there is also an interesting
relationship between the impossibility for a material body to reach
the velocity of light $v=1$ and the unattainability of the absolute
zero of temperature \cite{extremalbtz}.

It is also worth remembering that the property of $v<1$ was found for
near-horizon particle orbits in the background of near-extremal black holes
for the Kerr metric \cite{72} and in \cite{nh} for a much more general case.
Thus, we see an interesting analogy between limiting behaviors of
self-gravitating shells in
(2+1)-dimensional spacetimes and test particles in (3+1)-dimensional
spacetimes.

\subsection{Temperature  in the three horizon limits}

In the variables $\varepsilon$ and $\delta$ of Eqs.~(\ref{e}) and (\ref{d}),
the shell's local temperature $T$ in Eq.~\eqref{ne2} can be written as
\begin{equation}
T(\varepsilon,\delta,R)
=\frac{\delta ^{2}-\varepsilon ^{2}}{
2\pi \ell \delta \varepsilon \sqrt{1-\varepsilon ^{2}}}\,.
\label{tloc}
\end{equation}
For the \textit{Cases} 1-3,
the limits defined in Eqs.~(\ref{de1})-(\ref{de2})
yield from Eq.~\eqref{tloc} the expressions
for the local temperature as below.

\vskip 0.3cm \noindent \textit{Case 1:} For $\delta ={O}(1)$ and
$\varepsilon \to 0$,
\begin{equation}  \label{tdiv}
T(\varepsilon,\delta,r_+)= \frac{\delta }{2\pi \ell \varepsilon }\,,
\end{equation}
up to leading order.
Equation~(\ref{tdiv}) means that the temperature is divergent as $1/\varepsilon$.

\vskip 0.3cm \noindent \textit{Case 2:}
For $\delta =\frac{\varepsilon }{\lambda}$
and $\varepsilon \to0$,
\begin{equation}
T(\varepsilon,\delta,r_+) = \frac{1-\lambda ^{2}}{2\pi \ell \lambda }\,,
\label{t3}
\end{equation}
up to leading order.
Equation~(\ref{t3}) means that the local temperature is nonzero since
$\lambda$ is finite and fixed with
$\lambda<1$.
It is worth noting a simple formula that
follows from (\ref{p3}) and
(\ref{t3}) and relates the pressure and temperature in this
horizon limit, namely, 
$
\frac{p}{T}= \frac{1}{4G}\frac{1+\lambda^2-\lambda }{1-\lambda^2}
$.
Thus, if we believe that the
horizon of a black hole 
probes quantum gravity physics,
we find that in this case the
quantum gravity regime obeys an ideal gas law.

\vskip 0.3cm \noindent \textit{Case 3:} For $\delta=\varepsilon$
and $\varepsilon \to 0$,
\begin{equation}
T(\varepsilon,\delta,r_+) = {\rm finite}\,.
\label{tsingle}
\end{equation}
This was shown in \cite{extremalbtz}.  Equation~(\ref{tsingle}) does not
follow directly from Eq.~(\ref{tloc}).  The condition for $T$ should
be modified. It turns out that $T_0 $ may depend not only on $r_+$ and
$r_-$ but also on $R$. As a result, it may happen that $T_0\to 0$ but
the local temperature on the shell $T$ remains finite
\cite{extremalbtz}.

\section{Entropy: The three extremal BTZ black hole limits}
\label{sec5} 

Having carefully studied
the equations of state
for $p$, $v$, and $T$, 
we can now calculate the entropy 
by integrating the first law, see Eq.~\eqref{1st},
in all three cases.

\vskip 0.3cm \noindent {\it Case 1:}
$\delta ={O}(1)$ and $\varepsilon \to 0$.

\noindent
Here, 
we use first
the expressions in terms of $(\varepsilon,\delta,R)$,
i.e., Eqs.~\eqref{med}, \eqref{jed}, \eqref{pd}, \eqref{cd}, and \eqref{tloc}.
Then, one finds that the first law Eq.~\eqref{1st} can be expressed 
in terms of the differentials of $d\epsilon$,  $d\delta$, and  $dR$ as
$dS(\varepsilon,\delta,R)
=
\frac{\pi}{2G}
\Big(
-\frac{R \varepsilon}{\sqrt{1-\varepsilon^2}}
d\varepsilon+ \sqrt{1-\varepsilon^2}dR
\Big)$.
Then, taking $\varepsilon \to 0$, i.e., $R\to r_+$,
we get $dS(\varepsilon,\delta,r_+)
=
\frac{\pi}{2G}\,dr_+$. 
Since it does not depend on $\delta$ the expression is also
valid in the $\delta\to0$ case, i.e., in the extremal
case $r_+\to r_-$. Then integrating 
with the condition $S\to 0$ as $r_+\to 0$,
we get
in this extremal limit
\begin{equation}
S
=\frac{A_+}{4G} \,,
\label{sbh01}
\end{equation}
where $A_+=2\pi r_+$ is the area, i.e., the perimeter, of the shell,
i.e., the ring, see Eq.~\eqref{areah},
when it is pushed to its gravitational radius.
The
entropy in Eq.~\eqref{sbh01} 
is nothing but the Bekenstein-Hawking entropy, see
Eq.~\eqref{ent1bh}.

\vskip 0.3cm \noindent \textit{Case 2:}
$\delta =\frac{\varepsilon }{\lambda}$
and $\varepsilon \to0$.

\noindent
Here, we also have to use 
the expressions in terms of $(\varepsilon,\delta,R)$ 
i.e., Eqs.~\eqref{med}, \eqref{jed}, \eqref{pd}, \eqref{cd} and \eqref{tloc},
and then the first law Eq.~\eqref{1st} can be expressed 
in terms of the differentials of $d\epsilon$, $d\delta$, and $dR$, as
$dS(\varepsilon,\delta,R)
=
\frac{\pi}{2G}
\Big(
-\frac{R \varepsilon}{\sqrt{1-\varepsilon^2}}d\varepsilon
+
\sqrt{1-\varepsilon^2}dR
\Big)$, which is the same formula as in \textit{Case 1}.
Then, taking $\varepsilon \to 0$, i.e., $R\to r_+\to r_-$,
we get $dS(\varepsilon,\delta,r_+)
=
\frac{\pi}{2G}\,dr_+$.
This means that the entropy is independent of the parameter 
$\lambda$.
Then, integrating
with the condition $S\to 0$ as $r_+\to 0$,
we get
in this extremal limit
\begin{equation}
S
=\frac{A_+}{4G} \,,
\label{sbh02}
\end{equation}
where again $A_+=2\pi r_+$, see Eq.~\eqref{areah}. 
The entropy in Eq.~\eqref{sbh02} 
is again the Bekenstein-Hawking entropy,
see Eq.~\eqref{ent1bh}.
This result was not involved in former studies
\cite{btzshell,extremalbtz}.

\vskip 0.3cm \noindent \textit{Case 3:}
$\delta =\varepsilon$ and $\varepsilon \to0$.

\noindent
This case is special. 
One takes the extremality condition
$\delta=\varepsilon$ from the beginning
and thus another route to calculate the
entropy has to be followed.
This was performed in \cite{extremalbtz},
and the result is
\begin{equation}
S=S(A_+)\,,
\label{sbh03}
\end{equation}
where $S(A_+)$ is a well-behaved, but otherwise arbitrary, function of
$A_+$, see also Eq.~\eqref{ente11bh}.  One can argue, as was done in
\cite{extremalbtz}, that the lower and upper bounds for the entropy in
this case are given by the zero entropy, Eq.~\eqref{ent1ebh}, and the
Bekenstein-Hawking entropy, Eq.~\eqref{ent1bh}, i.e., $ 0\leq S(r_+)
\leq \frac{A_+}{4G}$, see Eq.~\eqref{lowupent1bh}.
In addition, 
Eq.~\eqref{sbh03} suggests that the entropy of an extremal black hole
does not take a unique value, but instead it may depend on the preceding
history that led to the
formation of precisely that extremal black hole,
see also \cite{pretisrvol}.

\section{Contributions to the
entropy in the three extremal horizon limits}
\label{sec6}

Finally, 
for all three different cases, 
we state
which terms in the first law \eqref{1st} give the dominant contributions to the entropy.

\vskip 0.3cm \noindent \textit{Case 1:} 
$\delta ={O}(1)$ and $\varepsilon \to 0$.

\noindent
We have that the
pressure term alone, see Eq.~(\ref{pdiv}), contributes to
the entropy.
Taking then
into account Eq.~(\ref{tdiv}), we obtain the Bekenstein-Hawking entropy (\ref{sbh01}).

\vskip 0.3cm \noindent \textit{Case 2:} $\delta =\frac{\varepsilon }{\lambda}$
and $\varepsilon \to
0 $.

\noindent
All three terms in the first law \eqref{1st} equally contribute to the entropy.
Thus, the mass, pressure, and circular velocity terms give contributions to
the Bekenstein-Hawking entropy \eqref{sbh02}.

\vskip 0.3cm \noindent \textit{Case 3:} $\delta=\varepsilon$ and $\varepsilon \to 0$.

\noindent
All three terms in the first law \eqref{1st} contribute to
the entropy, see \cite{extremalbtz}. 
We note that in contrast to the electrically charged case \cite{extremalshell}
the pressure does not vanish in the extremal limit and contributes
to the entropy in the first law as all other terms do, see Eq.~\eqref{sbh03}.

\vskip 0.3cm 
We summarize these results in Table I.

\begin{widetext}
\hskip -0.5cm
\begin{tabular}
[c]{|l|l|l|l|l|l|}\hline
Case & Pressure $p$ & Velocity  $v$ & Local temperature $T$ & Entropy $S(A_+)$ &
Contribution \\\hline
1 & Infinite & 1 & Infinite & $\frac{A_+}{4G}$~Eq.~\eqref{sbh01}  &
Pressure\\\hline
2 & Finite nonzero &$<1$ & Finite nonzero & $\frac{A_+}{4G}$~Eq.~\eqref{sbh02} & Mass, Pressure and
Angular velocity
\\\hline
3 & Finite nonzero & $\leq1$ & Finite zero and nonzero & $0\leq S(A_+)\leq \frac{A_+}{4G}$~Eq.~\eqref{sbh03}& 
Mass, Pressure and Angular velocity
\\\hline
\end{tabular}
\vskip 0.2cm
\noindent
Table 1. The contributions of the pressure
$p$, angular velocity $v$, and temperature $T$
to the entropy of the extremal black hole $S(A_+)$, according to the first law.
\label{tabent}
\end{widetext}

\section{Conclusions\label{sec7}}

We have presented a unified framework to explain how the different
entropies of an extremal BTZ black hole arise from an extremal shell.
{\it Case 1} and {\it Case 2} agree in the entropy but disagree in all
other thermodynamic quantities.  {\it Case 2} and {\it Case 3}
disagree in the entropy but agree in all other thermodynamic
quantities.  Therefore, in this sense {\it Case 2} is intermediate
between {\it Case 1} and {\it Case 3}.  These results complement the
former studies in a (2+1)-dimensional BTZ spacetime
\cite{quintalemosbtzshell,energycond,btzshell,extremalbtz} and have
much in common with those in the (3+1)-dimensional electrically
charged case \cite{charged,extremalshell}, in particular, with
\cite{lqzn}.

Consideration of astrophysically relevant rotating black holes in
(3+1) dimensions is too complex. In this regard, using the
(2+1)-dimensional rotating BTZ exact solution enables one to trace
quite subtle details that are expected in the more realistic (3+1)
case. Therefore, we hope that the present work can shed light on the
entropy issue for the (3+1)-dimensional black holes as well.


\section*{ACKNOWLEDGEMENTS}

We thank Funda\c c\~ao para a Ci\^encia e Tecnologia
(FCT), Portugal, for financial support through
Grant~No.~UID/FIS/00099/2013.~JPSL thanks 
Coordena\c c\~ao de Aperfei\c coamento do Pessoal de
N\'\i vel Superior (CAPES),
Brazil, for support within the Programa CSF-PVE,
Grant No.~88887.068694/2014-00.
JPSL also thanks an
FCT grant,
No.~SFRH/BSAB/128455/2017.~MM~thanks~FCT, 
for financial support through
Grant No.~SFRH/BPD/88299/2012.
OBZ thanks support from SFFR, Ukraine, Project No.~32367.
OBZ has also been partially supported by the Kazan
Federal University through a state grant for scientific
activities.


\begin{thebibliography}{99}

\bibitem{quintalemosbtzshell} J. P. S. Lemos and G. M. Quinta,
``Entropy of thin shells in a (2+1)-dimensional asymptotically AdS
spacetime and the BTZ black hole limit'', Phys. Rev. D {\bf 89},
084051 (2014); arXiv:1403.0579 [gr-qc].

\bibitem{energycond} J.~P.~S.~Lemos, F.~J.~Lopes, and M.~Minamitsuji,
``Rotating thin shells in (2+1)-dimensional asymptotically AdS
spacetimes: Mechanical properties, Machian effects, and energy
conditions'', Int. J.  Mod. Phys. D \textbf{24}, 1542022 (2015);
arXiv:1506.05454 [gr-qc].

\bibitem{btzshell} J.~P.~S.~Lemos, F.~J.~Lopes, M.~Minamitsuji, and
J.~V.~Rocha, ``Thermodynamics of rotating thin shells in the BTZ
spacetime'', Phys. Rev. D \textbf{92}, 064012 (2015); arXiv:1508.03642
[hep-th].

\bibitem{extremalbtz} J.~P.~S.~Lemos, M.~Minamitsuji, and
O.~B.~Zaslavskii, ``Thermodynamics of extremal rotating thin shells in
an extremal BTZ spacetime and the extremal black hole entropy'',
Phys. Rev. D \textbf{95}, 044003 (2017); arXiv:1701.02348 [hep-th].

\bibitem{charged} J.~P.~S.~Lemos, G.~M.~Quinta, and O.~B.~Zaslavskii,
``Entropy of a self-gravitating electrically charged thin shell and
the black hole limit'', Phys. Rev. D \textbf{91}, 104027 (2015);
arXiv:1503.00018 [hep-th].

\bibitem{extremalshell} J.~P.~S.~Lemos, G.~M.~Quinta, and
O.~B.~Zaslavskii, ``Entropy of an extremal electrically charged thin
shell and the extremal black hole'', Phys. Lett. B \textbf{750}, 306
(2015); arXiv:1505.05875 [hep-th].

\bibitem{lqzn} J. P. S. Lemos, G. M. Quinta, and O. B. Zaslavskii,
``Entropy of extremal black holes: Horizon limits through charged thin
shells in a unified approach'', Phys. Rev. D \textbf{93}, 084008
(2016); arXiv:1603.01628 [hep-th].

\bibitem{martin} E.~A.~Martinez, ``Fundamental thermodynamical
equation of a self-gravitating system'', Phys. Rev. D \textbf{53},
7062 (1996); arXiv:gr-qc/9601037.

\bibitem{quasi_bh1} J.~P.~S.~Lemos and O.~B.~Zaslavskii, ``Entropy of
quasiblack holes'', Phys. Rev. D \textbf{81}, 064012 (2010);
arXiv:arXiv:0904.1741 [gr-qc].

\bibitem{quasi_bh2} J.~P.~S.~Lemos and O.~B.~Zaslavskii, ``Entropy of
extremal black holes from entropy of quasiblack holes'', Phys. Lett. B
\textbf{695}, 37 (2011); arXiv:1011.2768 [gr-qc].

\bibitem{pretisrvol} F.~Pretorius, D.~Vollick, and W.~Israel, ``An
operational approach to black hole entropy'', Phys. Rev. D
\textbf{57}, 6311 (1998); arXiv:gr-qc/9712085.

\bibitem{btz} M.~Ba\~nados, C.~Teitelboim, and J.~Zanelli, ``The black
hole in three-dimensional space-time'', Phys. Rev. Lett. \textbf{69},
1849 (1992).

\bibitem{carlip} S.~Carlip, ``The (2+1)-dimensional black hole'',
Classical Quantum Gravity \textbf{12}, 2853 (1995); arXiv:gr-qc/9506079.

\bibitem{bek1} J.~D.~Bekenstein, ``Black holes and entropy'',
Phys. Rev. D \textbf{7}, 2333 (1973).

\bibitem{bch} J.~M.~Bardeen, B.~Carter, and S.~W.~Hawking, ``The four
laws of black hole mechanics'', Commun. Math. Phys. \textbf{31}, 161
(1973).

\bibitem{haw} S.~W.~Hawking, ``Particle creation by black holes'',
Commun.  Math. Phys. \textbf{43}, 199 (1975).

\bibitem{ebh2} C.~Teitelboim, ``Action and entropy of extreme and
nonextreme black holes'', Phys. Rev. D {\bf 51}, 4315 (1995);
arXiv:hep-th/9410103.

\bibitem{ebh1} S.~W.~Hawking, G.~T.~Horowitz, and S.~F.~Ross,
``Entropy, area, and black hole pairs'', Phys. Rev. D {\bf 51}, 4302
(1995); arXiv:gr-qc/9409013.

\bibitem{birmsacsen} D. Birmingham, I. Sachs, and S. Sen, ``Entropy of
three-dimensional black holes in string theory'',
Phys. Lett. B \textbf{424}, 275 (1998); arXiv:hep-th/9801019.

\bibitem{string11} C. V. Johnson, R. R. Khuri, and R. C. Myers,
``Entropy of 4-D extremal black holes'', Phys. Lett. B \textbf{378}, 78
(1996); arXiv:hep-th/9603061.

\bibitem{string1} A.~Strominger and C.~Vafa, ``Microscopic origin of
the Bekenstein-Hawking entropy'', Phys. Lett. B {\bf 379}, 99 (1996);
arXiv:hep-th/9601029.

\bibitem{string2} A.~Sen, ``Microscopic and macroscopic entropy of
extremal black holes in string theory'', General Relativ. Gravit.  {\bf 46},
1711 (2014); arXiv:1402.0109 [hep-th].

\bibitem{ebh3} G.~W.~Gibbons and R.~E.~Kallosh, ``Topology, entropy and
Witten index of dilaton black holes'', Phys. Rev. D \textbf{51}, 2839
(1995); arXiv:hep-th/9407118.

\bibitem{ebh4} A.~Ghosh and P.~Mitra, ``Understanding the area proposal for
extremal black hole entropy'', Phys. Rev. Lett. \textbf{78}, 1858 (1997);
arXiv:hep-th/9609006.

\bibitem{ebh5} S.~Hod, ``Evidence for a null entropy of extremal black
holes'', Phys. Rev. D \textbf{61}, 084018 (2000); arXiv:gr-qc/0004003.

\bibitem{ebh6} S.~M.~Carroll, M.~C.~Johnson, and L.~Randall,
``Extremal limits and black hole entropy'', J.  High Energy
Phys. \textbf{11} (2009) 109; arXiv:0901.0931 [hep-th].

\bibitem{ebh7} A.~Edery and B.~Constantineau, ``Extremal black holes,
gravitational entropy and nonstationary metric fields'', Classical
Quantum
Gravity \textbf{28}, 045003 (2011); arXiv:1010.5844 [gr-qc].

\bibitem{ghoshmitra} A.~Ghosh and P.~Mitra, ``Entropy for extremal
Reissner-Nordstr\"om black holes'', Phys. Lett. B \textbf{357}, 295 (1995); 
arXiv:hep-th/9411128.

\bibitem{string3} R.~B.~Mann and S.~N.~Solodukhin, ``Universality of quantum
entropy for extreme black holes'', Nucl. Phys. \textbf{B523}, 293 (1998);
arXiv:hep-th/9709064.

\bibitem{string4} C.~Kiefer and J.~Louko, ``Hamiltonian evolution and
quantization for extremal black holes'', Annalen Phys. \textbf{8}, 67
(1999); arXiv:gr-qc/9809005.

\bibitem{string5} G.~A.~S.~Dias and J.~P.~S.~Lemos, ``Hamiltonian
thermodynamics of $d$-dimensional ($d \geq 4$) Reissner-Nordstr\"om anti-de
Sitter black holes with spherical, planar, and hyperbolic topology'', Phys.\
Rev. D \textbf{79}, 044013 (2009); arXiv:0901.0278 [gr-qc].

\bibitem{cano1} O.~B.~Zaslavskii, ``Entropy of quantum fields for nonextreme
black holes in the extreme limit'', Phys. Rev. D \textbf{57}, 6265 (1998):
arXiv:gr-qc/9708027.


\bibitem{72} J. M. Bardeen, W. H. Press, and S. A. Teukolsky, ``Rotating black
holes: Locally nonrotating frames, energy extraction, and scalar synchrotron
radiation'', Astrophys. J. \textbf{178}, 347 (1972).

\bibitem{nh} O. B. Zaslavskii, ``Near-horizon circular orbits and extremal limit
for dirty rotating black holes'', Phys. Rev. D \textbf{92}, 044017 (2015);
arXiv:1506.00148 [gr-qc].




\end{thebibliography}
\end{document}